\documentstyle [12pt,epsfig]{article} 
\makeatletter

\@addtoreset{equation}{section}
\makeatother

\newcommand{\ba}{\begin{eqnarray}}
\newcommand{\ea}{\end{eqnarray}}

\newcommand{\RA}{{\rm A}}
\newcommand{\RB}{{\rm B}}
\newcommand{\RC}{{\rm C}}
\newcommand{\RR}{{\rm R}}
\newcommand{\RF}{{\rm R}}
\newcommand{\tF}{t_{\rm F}}
\newcommand{\tFP}{t_{\rm F'}}

\begin{document}
\newcommand{\BS}{\bigskip}
\newcommand{\SECTION}[1]{\BS{\large\section{\bf #1}}}
\newcommand{\SUBSECTION}[1]{\BS{\large\subsection{\bf #1}}}
\newcommand{\SUBSUBSECTION}[1]{\BS{\large\subsubsection{\bf #1}}}

\begin{titlepage}
\begin{center}
\vspace*{2cm}
{\large \bf Space-time attributes of physical objects and the laws of
   space-time physics}  
\vspace*{1.5cm}
\end{center}
\begin{center}
{\bf J.H.Field }
\end{center}
\begin{center}
{ 
D\'{e}partement de Physique Nucl\'{e}aire et Corpusculaire
 Universit\'{e} de Gen\`{e}ve . 24, quai Ernest-Ansermet
 CH-1211 Gen\`{e}ve 4.
}
\newline
\newline
   E-mail: john.field@cern.ch
\end{center}
\vspace*{2cm}
\begin{abstract}
   Physical time intervals are attributes of single physical object whereas physical space
   intervals are a relational attribute of two physical objects. Some consequences of
   the breaking of the space-time exchange symmetry inherent in the Lorentz transformation
   following from the above distinction are investigated. In particular, it is shown that
   the relativity of simultaneity and length contraction effects which naively follow
   from space-time symmetry of the Lorentz transformation do not occur. Seven laws
   describing the relation between observations of space intervals, time intervals and
   velocities in different reference frames are given. Only two of these laws are respected
   by conventional special relativity theory.

\vspace*{1cm}
\end{abstract}
\end{titlepage}
 
\SECTION{\bf{Introduction}}
        The purpose of the present paper is an in-depth discussion of the operational
    physical meanings of the symbols in the Galilean equation of motion of a physical
    object:
      \begin{equation}
       \Delta x = v \Delta t
      \end{equation}
      relating space intervals, $\Delta x$, time intervals, $\Delta t$, and velocity, $v$.
      The general case, for small values of $\Delta x$ and  $\Delta t$, where  $v$ is
       time-dependent, as well as uniform motion where $v = {\rm constant}$ will be 
        considered. The specific question that is addressed is the attribution of the symbols
   $\Delta x$, $\Delta t$ and $v$ to one or more discrete physical objects A,B,... . The
    symbol C is used to denote a clock, which, as discussed below, is the class of
    self-age-recording objects.
     \par The plan of the paper is as follows. The following two sections discuss the definitions
     of physical time intervals and space intervals respectively, and how they may be measured.
      In Section 4 the different concepts of frame velocity and relative velocity are defined.
      Section 5 considers the initial configuration of the `twin paradox' thought experiment,
      and shows by application of the Principle of Sufficient Reason (or alternatively, by
      {\it reductio ad absurdum}) the invariance of a length interval measured in two different inertial 
       frames. After recalling the important concepts of `primary' and `reciprocal' space-time
     experiments and `base' and `travelling' frames, as introduced in Refs.~\cite{JHFSTP3,JHFRECP}, 
     Section 6 analyses the different time dilatation effects in primary and reciprocal experiments.
     The formulae obtained demonstrate the spurious nature of the `relativity of simultaneity' effect
     of standard special relativity theory (SSRT). In Section 7, transformation formulae, between two
      inertial frames, of the relative velocity of two objects are obtained. In the concluding
     Section 8, the conclusions of the paper are summarised in seven laws of space-time physics, five
     of which are not respected by the predictions of SSRT.
  
\SECTION{\bf{Temporal Intervals, $\Delta t$}}
  The crucial point to be noticed here is that, in general, a time interval
  (unlike a space interval) is {\it an attribute of a single physical object}. This concept
  is a commonsense one familiar in the guise of the age of any living creature or plant,
   which is the time interval between the present and the birth date of any animal or the
   germination date of any plant\footnote{It would be more consistent to specify the age of an animal
   as the time interval between conception and the present, but this is not the social convention.}.
    \par In order to make quantitative physical statements concerning age or time intervals,
    it is necessary to introduce the concept of a clock, which is defined as {\it a physical object
    that records an interval of its own age}. In order to do this the object must furnish, at any
    instant, 1, a number $t_1$, conventionally termed the {\it epoch} of the instant. The time
    interval between instants 1 and 2, is then defined as the difference of the epochs of 1 and 2:
     \begin{equation}
       \Delta t_{21} \equiv t_2- t_1
    \end{equation}
      A good example of a self-age-recording object is a radioactive clock, as suggested by
      Langevin~\cite{Langevin}. This is a ponderable object (a radioactive `source') that is
    doped at epoch $t_0$ with $N_0$ atoms of a radioactive substance of known mean decay time $\tau_D$.
    In virtue of the radioactive decay law, the age, $a_1$, of the source at epoch $t_1$ when
     $N_1$ atoms remain undecayed is:
  \begin{equation}
    a_1 \equiv  t_1- t_0 = \tau_D \ln\left(\frac{N_0}{N_1}\right)
  \end{equation}
   As pointed out by Langevin, a practical way to measure  $a_1$ (since it is evidently not possible
    to count $N_1$, even if the value of $N_0$ is experimentally controllable by chemical means)
      is to place the source in the same detector at the different epochs $t_0$ and $t_1$ and measure
     the corresponding activities $A_0$ and $A_1$ (the numbers of disintegrations per
      unit of time recorded). The age of the source is then given by the relation
      \begin{equation}
    a_1 = \tau_D \ln\left(\frac{A_0}{A_1}\right)
  \end{equation}
    Evidently, because of the statistical nature of the decay process, the age determined by (2.3) 
   is subject to a statistical uncertainty that depends on the value of $N_0$ and the efficiency
    of the detector. The `true' value, still, however, dependent of the experimental uncertainty
     in $\tau_D$, is given by (2.2) or (2.3) in the limit $N_0 \rightarrow \infty$.  Consideration of
      such radioactive clocks makes quite transparent the absurdity of certain predictions
      of SSRT~\cite{JHFSTP3,JHFLLT}.
     \par More conventional clocks are based on periodic physical processes such as the orbital period
     of a planet, the oscillation period of a pendulum, balance wheel or quartz crystal, or the
     frequency, $\nu_{\gamma}$, associated via the Planck-Einstein relation $E_{\gamma} = h \nu_{\gamma}$ with the
     energy of a photon $E_{\gamma}$ emitted in a transition between two well-defined atomic
     energy levels. There is therefore an irreducible connection between the concept of a physical
     time interval and the time dependence or frequency (exponential decay law, repeated periodic motion,
     or the proportionality of the energy and frequency of photons in quantum mechanics) of definite
     physical processes. This means that there is no conceptual ambiguity in the operational definition
     of a physical time interval.
       \par Although time, as specified by the epoch number $t({\rm C})$ of a clock C, is a fundamental
        attribute of C, it can be extended to every member of an array of clocks C$_1$,C$_2$,.. which
       are at rest  relative to each other, by introducing the further concepts of a
       {\it reference frame} F, and {\it frame time} $\tF$. The frame F contains a system of
        spatial coordinates that specifies the relative spatial positions of C$_1$,C$_2$,.., and is the
      the common proper frame of any of the clocks, which may be in arbitary (uniform or accelerated) motion
   relative to an inertial observer. If, at any instant in the frame F, the epochs recorded
    by C$_1$,C$_2$,.. are the same:
      \begin{equation}
       t({\rm C_1}) = t({\rm C_2}) = ... \equiv  \tF
      \end{equation}
        the array of clocks is said to be {\it synchronised}\footnote{To be contrasted with
           a {\it synchronous} array of clocks which run at the same speed but may indicate
               different epochs.}. In the case that F is an inertial frame,
           the clocks may be mutually synchronised using Einstein's well known light-signal 
          procedure~\cite{Ein1} that relies on the assumption of the isotropy of the
         speed of light. More general synchronisation  procedures, not using light signals in
      free space, and applicable also to accelerated frames, are described in Ref.~\cite{JHFSTP1}.
       \par Of great importance for the following discussion is the concept of a 
         {\it corresponding epoch} in two different reference frames F and F'. This is
       defined as follows:
       \par {\it A corresponding epoch of two frames} F {\it and} F' {\it  in relative motion is one for which one or
          more objects at rest in F are seen to be spatially contiguous with one or more objects in
           at rest in F' by observers in both frames }.
       \par In the notation for space intervals to be introduced in the following section,
         if
   \begin{equation}
      \Delta x(\RA,\RA',\tF) = \Delta x'(\RA,\RA',\tFP) = 0
   \end{equation}   
      then $\tF$ and  $\tFP$ are {\it corresponding epochs}. If it happens that 
       $\tF =,\tFP$ in Eq. (2.5) then {\it clocks in the frames} F {\it and} F' {\it are said to be
       synchronised}. In the case that F and F' are inertial
      frames S and S' and A and A'are placed at the origins of spatial coordinates in S and S'
      respectively, the condition (2.5) is identical to Einstein's definition of local clock
        synchronisation
      between two inertial frames~\cite{Ein1}. Eq. (2.5) also corresponds to the `system external'
      synchronisation of clocks in different reference frames of Mansouri and Sexl~\cite{MS}.

 \SECTION{\bf{Spatial Intervals, $\Delta x$}}
       In the following, only one-dimensional spatial separations and motion are considered.
       Because of the assumed isotropy of three-dimensional space no loss of generality
       is occasioned for the problems considered in the present paper.
        \par In contrast to a time interval, a spatial interval is {\it an attribute of two distinct
        physical objects}\footnote{In the case when the spatial interval corresponds to a dimension
       of an extended physical object, it is necessary to introduce localised portions of
       the object situated at the ends, that play the role of two discrete separated objects. In one
       dimension, any extended object has two and only two `ends', the distance between which
         defines the size of the object in this dimension. In the case of redundant definition
        of the ends (e.g. a rectangular bar) any two localised `end objects' can be used to
        specify the length of the bar.}. It follows from this that a spatial interval at a
       given epoch $\tF$ in a frame F requires two labels A and B corresponding to the
      related physical objects in its definition: $\Delta x(\RA,\RB,\tF)$, where the
      spatial coordinate $x$ is defined in the frame F, in contrast to
      a time interval which requires only one label to specify it: $\Delta t(\RA)$, for
       a particular object, A, or $\tF$ for a particular frame, F. It is this difference of the attribution
      of temporal and spatial intervals to physical objects that breaks the mathematical
      space-time exchange symmetry~\cite{JHFAJP2} of the Lorentz transformation equations\footnote{Introducing a
       temporal coordinate with dimension [L] according to $x_0 = ct$, the space-time Lorentz
        transformation equations
       are written as $x' = \gamma(x-\beta x_0)$, $x_0' = \gamma(x_0-\beta x)$. This pair of equations
       is invariant under exchange of space-time coordinates: $x \leftrightarrow x_0$,
     $x' \leftrightarrow x'_0$.}
     in their application to actual physical problems involving the description of synchronised 
     clocks.
      \par For a particular spatial coordinate, $x$, in the frame F, the definition of the spatial
      separation of A and B is
  \begin{equation}
      \Delta x(\RA,\RA',\tF) \equiv x(\RA,\tF)-x(\RB,\tF)
   \end{equation} 
      The interval $ \Delta x$ is invariant with respect to different choices of coordinate origin,
         equivalent to the coordinate transformation: $x \rightarrow x +C$, where $C$ is an arbitary
    constant. 
    \par The measured spatial separation of two objects in arbitary motion in a reference frame F
         is defined as follows:
     \par{\it If two objects} A' {\it and} B' {\it in arbitary motion in a reference frame} F
       {\it are spatially contiguous with the objects}  A {\it and} B {\it at rest in} F {\it at epoch} $\tF$,
        {\it the spatial separation of} A' {\it and} B' {\it at this epoch is the same as that of}  A {\it and} B.
    \par This is just a statement of the methodology of `ruler measurement' of the separation of two
       objects at a given epoch, as discussed in detail in Ref.~\cite{JHFSTP1}. The expression
        `spatially contiguous' in the above definition is equivalent, in the present context, to 
         `have the same $x$-coordinate'. As must be the case for non-colliding physical objects of
     non-vanishing
          lateral dimensions, the $y$ and/or $z$ coordinates of the objects may be different.

  \SECTION{\bf{Frame velocity, $v$, and Relative Velocity, $u$}}
     The {\it frame velocity}, $v_{\RA'}$ is an attribute of a single physical object, A', and depends on the frame 
     of reference, F, in which it is defined. If the frame F is understood in the symbol $v$ (so that, for example,
      the frame velocity in the frame F' is denoted by $v'$) $v_{\RA'}$ depends only on the label A'
       $v_{\RA'} \equiv v(\RA',\tF)$.
       Other objects B',C',... at rest in the proper frame F' of A', but at different spatial locations,
      have the same frame velocity as A' in F at any epoch:
      \begin{equation}         
    v(\RA',\tF) =  v(\RB',\tF) =  v(\RC',\tF) = ...= v(\tF) 
       \end{equation}
     If A is an object at rest at an arbitary position in F, the definition of $v(\tF)$ in terms if the epoch $\tF$
      and the spatial interval $\Delta x$ introduced above is:
    \begin{equation} 
     v(\tF) \equiv \frac{d[\Delta x(\RA',\RA,\tF)]}{d \tF} = \frac{d[x(\RA',\tF)-x(\RA,\tF) ]}{d \tF}
       =  \frac{dx(\RA',\tF)}{d \tF} = \frac{dx(\RB',\tF)}{d \tF} = ...
    \end{equation}
        since $dx(\RA,\tF)/d \tF = 0$. The parameter $v$ (now independent of $\tF$) is that appearing in the
       space-time Lorentz transformation between the inertial frames S and S':
       \begin{eqnarray}
       x' & = & \gamma(v)[x(\tF)-v\tF] \\
       t' & = & \gamma(v)[\tF-\frac{vx(\tF)}{c^2}]
       \end{eqnarray}
     where $\gamma(v) \equiv 1/\sqrt{1-(v/c)^2}$, in which case $v =  v_{\RA'} = {\rm constant}$ 
       where A' is any object at rest in S'.
       \par The {\it relative velocity}, $u$, of two objects A', A'' in arbitary motion in the frame F, unlike $v$, 
       is a relational attribute of both A'and A''. It is defined in a similar manner to $v$ in Eq. (4.2):
     \begin{eqnarray} 
     u(\RA',\RA'',\tF) & \equiv & \frac{d[\Delta x(\RA',\RA'',\tF)]}{d \tF} = \frac{d[x(\RA',\tF)-x(\RA'',\tF) ]}{d \tF}
       \nonumber \\
      & = &   \frac{dx(\RA',\tF)}{d \tF}-  \frac{dx(\RA'',\tF)}{d \tF} = v(\RA',\tF) - v(\RA'',\tF)
    \end{eqnarray}
         The velocity of A' relative to A'' in F is thus equal the the difference of their frame
       velocities. Note that, although the frame velocity of a ponderable physical object has, in 
       SSRT, the upper limit of $c$, the velocity of light in free space, the upper limit on $u$ in the
       frame F, given by (4.5), is $2c$.
 \SECTION{\bf{Frame Invariance of Length Intervals}}
      For illustrative purposes, and to demonstrate the flaw in the standard text book interpretation
     of the `twin paradox' thought experiment~\cite{Langevin} the journey, at uniform frame velocity $v$, in the rest
      frame, S, of the Earth (E) of a space ship R$_1$ to a distant star Sirius (Si), assumed to be at rest in S,
      is considered.

\begin{figure}[htbp]
\begin{center}\hspace*{-0.5cm}\mbox{
\epsfysize15.0cm\epsffile{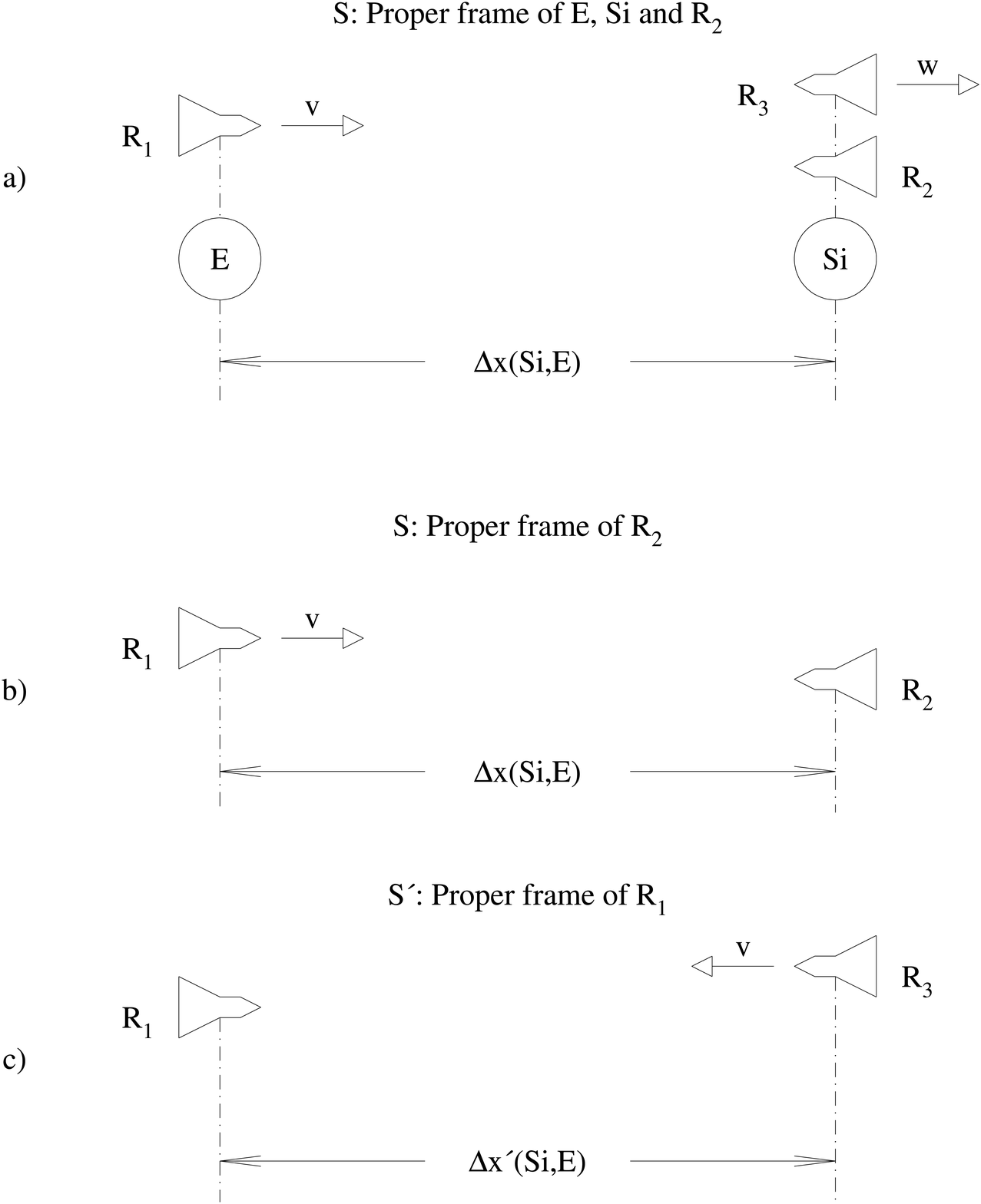}}
\caption{{\em Spatial and kinematical configurations of the Earth (E) the star Sirius (Si) and the spaceships
   R$_1$, R$_2$ and  R$_3$ at the beginning of the voyage of  R$_1$ to Sirius in the `twin paradox' thought
   experiment. a), b) in the rest frame, S, of E, Si and  R$_2$; c) in the rest frame, S', of
    R$_1$. The perfect symmetry of the configurations of R$_1$ and R$_2$ in b) and  R$_1$ and R$_3$ in c) togther
    with the Principle of Sufficient Reason requires that
     $\Delta x'({\rm Si} ,{\rm E}) =   \Delta x({\rm Si} ,{\rm E})$.}}
\label{fig1} 
\end{center}
\end{figure}

      \par At the start of the journey, when R$_1$ and the Earth are aligned, two other spaceships  R$_2$ and
       R$_3$ aligned with Sirius are introduced. R$_2$ is at rest in S while  R$_3$ has a certain frame velocity
       $w$, to be determined below, such that in the proper frame, S' of  R$_1$, R$_3$ has velocity $v$ 
       in the direction of the Earth. In Fig.~\ref{fig1}a is shown the disposition of objects in S at the start
       of the journey. The following space-interval equalities hold in S at the epoch shown in Fig.~\ref{fig1}a:
     \begin{equation}
       \Delta x({\rm Si} ,{\rm E}) = \Delta x({\rm R}_2 ,{\rm E}) =   \Delta x({\rm R}_3 ,{\rm E}) =
           \Delta x({\rm Si},{\rm R}_1) = \Delta x({\rm R}_2,{\rm R}_1) =  \Delta x({\rm R}_3,{\rm R}_1)
        \end{equation}
    Similarly, in the rest frame of  R$_1$ at the corresponding epoch:
   \begin{equation}
       \Delta x'({\rm Si} ,{\rm E}) = \Delta x'({\rm R}_2 ,{\rm E}) =   \Delta x'({\rm R}_3 ,{\rm E}) =
           \Delta x'({\rm Si},{\rm R}_1) = \Delta x'({\rm R}_2,{\rm R}_1) =  \Delta x'({\rm R}_3,{\rm R}_1)
        \end{equation}
     Attention is now fixed on the spatial configuration of  R$_1$ and  R$_2$ in S at this epoch (Fig.~\ref{fig1}b) and
      that of R$_1$ and  R$_3$ in S' at the corresponding epoch (Fig.~\ref{fig1}c).
        From Eq. (5.1) the separation of  R$_1$ and  R$_2$ in Fig. 1b is $\Delta x \equiv \Delta x({\rm Si} ,{\rm E})$
        while from Eq. (5.2) that of  R$_1$ and  R$_3$ in Fig.~\ref{fig1}c is $\Delta x' \equiv \Delta x'({\rm Si} ,{\rm E})$.
      \par Inspection of Fig.~\ref{fig1}b and Fig.~\ref{fig1}c reveals perfect symmetry between the two kinematical configurations.
       In fact they are related to each other by the parity operation ---reflection in the plane midway between
        the Earth and Sirius perpendicular to the directions of motion of  R$_1$ and R$_3$. In these circumstances
       what possible physical effect, depending only on the value of the velocity, $v$, could result in a {\it difference}
       between  $\Delta x$ and $\Delta x'$?  There is none. It therefore follows on application of Liebnitz'
       Principle of Sufficient Reason that $\Delta x = \Delta x'$. 
       \par The same conclusion is reached by applying to the configurations of Fig.~\ref{fig1}b and 1c the relativistic
       reciprocity relation\footnote{Pauli's reciprocity relation is a special case of the Measurement
        Reciprocity Postulate (MRP)~\cite{JHFSTP1,JHFHPA,JHFASIL} which is a simple, purely kinematical, form of the
        Special Relativity Principle from which the Lorentz transformation may be derived without consideration
        of light signals or classical electromagnetism~\cite{JHFSTP1,Ignatowsky,JHFHPA}}. used by Ignatowsky~\cite{Ignatowsky}
      to derive the Lorentz transformation, as cited
       by Pauli~\cite{Pauli}\footnote{Pauli's frame labels K and K' are replaced by S and S' for conformity
       with the notation of the present paper.}:
        \par{\tt The contraction of length at rest in S' and observed in S is equal to \newline that of lengths
          at rest in S and observed from S'.} 
         \par It the velocity-dependent contraction factor is denoted by $\alpha(v)$, the first condition
          applied to the configurations of Fig.~\ref{fig1}b and 1c gives:
       \begin{equation}
             \alpha(v)\Delta x' =\Delta x
    \end{equation}
             while the second gives:
       \begin{equation}
             \alpha(v)\Delta x =\Delta x'
    \end{equation}  
      Combining (5.3) and (5.4),
   \begin{equation}
   \alpha(v)\Delta x' =  \alpha(v)[\alpha(v)\Delta x] =  \alpha(v)^2 \Delta x =\Delta x
   \end{equation}
 So either $\Delta x' = \Delta x = 0$ or $\alpha(v)^2 = 1$. In the present case since $\Delta x$ and $\Delta x'$ are
 both non-zero and positive it follows that $\alpha(v) = 1$ so that
       \begin{equation}
         \Delta x' =\Delta x
    \end{equation}
       
   In the interpretation of the twin paradox in SSRT it is instead assumed that `length contraction' occurs
   in an asymmetric manner in the rest frame of  R$_1$ so that $\Delta x' =\Delta x/\gamma(v)$. 
     The erroneous nature of this prediction and the absurd consequences that
    it entails are discussed at length in Ref.~\cite{JHFSTP3}. If $\alpha(v)$ is identified with $\gamma(v)$, 
    Eqs. (5.3) and (5.4) require that $v = 0$, contradicting the initial assumption of Fig.\ref{fig1} that $v > 0$. If the
    reciprocity relation quoted by Pauli holds, the
     identification of the contraction factor with $\gamma(v)$ is thus excluded by
     {\it reductio ad absurdum}\footnote{This argument was previously given in Ref.~\cite{JHFFT}.}.
   \par The calculation above considered only spatial intervals in two different inertial frames, but
    the same invariance of spatial intervals at corresponding epochs is found for two frames
    each undergoing arbitary accelerated motion since the above reasoning is valid in the 
    comoving inertial frames at the corresponding epoch. See Ref.~\cite{JHFASIL} for a discussion
    of the spatial interval between two objects undergoing identical accelerated motion.
 \SECTION{\bf{Time Dilatation and Reciprocal Experiments}}
    This section specialises to the case of inertial frames S and S' in the standard configuration where S' moves 
    with velocity $v$ along the positive $x$-axis and the $x$- and $x'$-axes are parallel.
    The space-time Lorentz transformation (LT) gives a relation between time intervals $\Delta t$ in S and
 $\Delta t'$ in S' as registered by clocks at rest in these frames. Two physically independent and reciprocal
   experiments are possible~\cite{JHFSTP3,JHFRECP}. 
    In the first {\it primary} experiment a clock at rest in S' is
    observed from S and compared with a clock at rest in the latter frame. In the second {\it reciprocal} 
    experiment a clock at rest in S is observed from S' and compared with clock, like the observer, at rest
   in S'. The LT equations are completely different in these two {\it independent experiments}. In the primary
   experiment only the spatial coordinates of the clock at rest in S' appear, whereas in the reciprocal
   experiment only those of the clock at rest in S are used. In the primary experiment the position of the clock
   at rest in S is arbitary, while in the reciprocal experiment the position of the clock at rest in S' is arbitary,
   and the spatial coordinates of these clocks do not appear in the LT equations. 
    \par In the primary experiment an interval on the world line of a clock C' at rest at an arbitary position
    in S' is given by Eq. (1.1) as
    \begin{equation}
       \Delta x(\RC') = v \Delta t(\RC)_B
      \end{equation}
   where C is a clock at rest at an arbitary position in S. 
   The subscript $B$  on $\Delta t(\RC)_B$ stands for {\it base frame} and indicates that the time
    interval is recorded by a clock at rest in the same frame, S, as the the observer.
   The corresponding time interval in S' is given by the time LT as:
   \begin{equation}
   \Delta t'(\RC')_T = \gamma(v)\left[ \Delta t(\RC)_B-\frac{v \Delta x(\RC')}{c^2}\right] 
   \end{equation}
    The subscript $T$ on $\Delta t'(\RC')$ stands for {\it Travelling frame} indicating that the
    clock is in motion relative to the observer at rest in the base frame S. Thus, in the primary
     experiment, S' is the travelling frame. Combining (6.1) and (6.2) gives the Time Dilatation (TD) 
     relation  for the primary experiment:
    \begin{equation}  
     \Delta t(\RC)_B = \gamma(v)  \Delta t'(\RC')_T
   \end{equation}
      \par In the reciprocal experiment where, by definition, the clock C, at an arbitary position
     in S, moves with speed $v$ parallel to the negative $x'$-axis in S', an interval of the
      world line of C is given by Eq. (1.1) as:
      \begin{equation}
       \Delta x'(\RC) = -v \Delta t'(\RC')_B
      \end{equation}
      where the clock C' is at an arbitary position in S'. 
      Combining the appropriate time LT:
   \begin{equation}
   \Delta t(\RC)_T = \gamma(v)\left[ \Delta t'(\RC')_B+\frac{v \Delta x(\RC)}{c^2}\right] 
   \end{equation}
    with (6.4) gives the TD relation for the reciprocal experiment:
     \begin{equation}  
     \Delta t'(\RC')_B = \gamma(v)  \Delta t(\RC)_T
   \end{equation}
      Notice that since the quantities represented by the symbols $\Delta t(\RC)_B$, $\Delta t(\RC)_T$,
      $\Delta t'(\RC')_B$ and  $\Delta t'(\RC')_T$ are physically distinct there is no antinomy between
      Eqs. (6.3) and (6.6) and the condition $\gamma(v) > 1$, as there is between  Eqs. (5.3) and (5.4)
     and the `length contraction' condition $\alpha(v) = \gamma(v) > 1$~\footnote{It was the failure to notice that
      $\Delta t(\RC)_B$ is not the same physical quantity as $\Delta t(\RC)_T$ and $\Delta t'(\RC')_B$ is not
      the same physical quantity as $\Delta t'(\RC')_T$ that led Dingle to
       wrongly conclude, by a {\it reductio ad absurdum} argument, similar to that of Eqns(5.3) and (5.4) above,
      that SSRT was self-contradictory and therefore untenable~\cite{Dingle}.}. This shows clearly that the
      apparent
     space-time symmetry of the LT equations breaks down when they are applied to actual physical 
     problems and care is taken over the precise operational meaning of the symbols for
     spatial and temporal intervals that appear in the equations.
      \par Since the spatial positions of the clocks C and C' in the TD relations (6.3) and (6.6)
      are arbitary, it is an immediate consequence that clocks in S and S', once synchronised
      at a corresponding epoch, remain so at all later epochs. Suppose that $\RC'_1$ and  $\RC_1$ 
      are synchronised so that:
   \begin{equation}  
     t(\RC_1)_B = \gamma(v) t'(\RC'_1)_T
   \end{equation}
     which implies that $t'(\RC'_1)_T = 0$ when $t(\RC_1)_B = 0$, and that 
      $\RC'_2$ and  $\RC_2$ are similarly synchronised so that:
     \begin{equation}  
     t(\RC_2)_B = \gamma(v) t'(\RC'_2)_T
   \end{equation}
     If now, it so happens that $\RC_1$ and  $\RC_2$, at rest in S, were previously
      synchronised so that 
  \begin{equation} 
  t(\RC_1)_B = t(\RC_2)_B \equiv t_B
  \end{equation}
  it follows from (6.8) and (6.9) that      
    \begin{equation}  
     t'(\RC'_1)_T = t'(\RC'_2)_T \equiv t'_T = t_B/\gamma(v)  
   \end{equation}
   $\RC'_1$ and  $\RC'_2$ ---clocks at arbitary positions in S'--- are then also synchronised at any
     epoch ---there is no `relativity of simultaneity' effect.

 \SECTION{\bf{Relative Velocity Transformation Formulae}}
  A transparent way to derive the transformation formula, between two inertial frames, of the relative
  velocity of two objects in uniform motion is by consideration of the outward journey of R$_1$ from Earth 
  to Sirius in the twin paradox experiment shown in Fig. 1. In order to deduce the value of the velocity, $w$,
  of R$_3$ in Fig. 1a, it is convenient to consider a configuration where R$_1$ and  R$_3$ are aligned at 
  the beginning of the journey
 as shown in Fig. 2a. The velocity of R$_1$ relative to R$_3$ has the same value $v-w$ in Fig 1a and Fig. 2a. 
  \par The configuration of the objects at the end of the journey when $T \equiv t = \gamma(v)t' \equiv \gamma(v)T'$
  as given by the TD relation (6.3), are shown in the frame S (the proper frame of the Earth and Sirius) in Fig. 2b,
   and in S' (the proper frame of R$_1$) in Fig. 2c. Denoting the velocity of R$_3$ relative to R$_1$ in S' by
   $u'(\RR_3,\RR_1)$ and that of the Earth relative to R$_1$ in the same frame by $u'({\rm E},\RR_1)$, the geometry of Figs. 2a and
   2b gives the relations:
     \begin{eqnarray}
    T & = & \frac{\Delta x(\RR_1,{\rm E},T)}{v} =  \frac{\Delta x(\RR_1,\RR_3,T)}{v-w} \\ 
    T' & = & \frac{\Delta x'(\RR_1,{\rm E},T')}{u'({\rm E},\RR_1)} =  \frac{\Delta x'(\RR_1,\RR_3,T')}{u'(\RR_3,\RR_1)} 
   \end{eqnarray}
    Since 
  \begin{eqnarray}
    \Delta x'(\RR_1,{\rm E},T') & = & \Delta x(\RR_1,{\rm E},T) \\
    \Delta x'(\RR_1,\RR_3,T') & = & \Delta x(\RR_1,\RR_3,T) \\
     T & = & \gamma(v)T'
  \end{eqnarray} 
    (7.1) and (7.2) give the relative velocity transformation formulae:
  \begin{eqnarray}
 u'({\rm E},\RR_1) & = & \gamma(v) v = \gamma u(\RR_1,{\rm E}) \\
 u'(\RR_3,\RR_1) & = & \gamma(v)(v-w) = \gamma(v)u(\RR_1,\RR_3)
 \end{eqnarray} 
    Note that (7.6) is a special case of (7.7) when $w = 0$, so that R$_3$ is at rest relative to the Earth.
   The change in the order of the labels on the right and left sides of (7.6) and (7.7) corresponds
   to the change in direction of the relative velocity vector in the transformation from S to S'.

\begin{figure}[htbp]
\begin{center}\hspace*{-0.5cm}\mbox{
\epsfysize15.0cm\epsffile{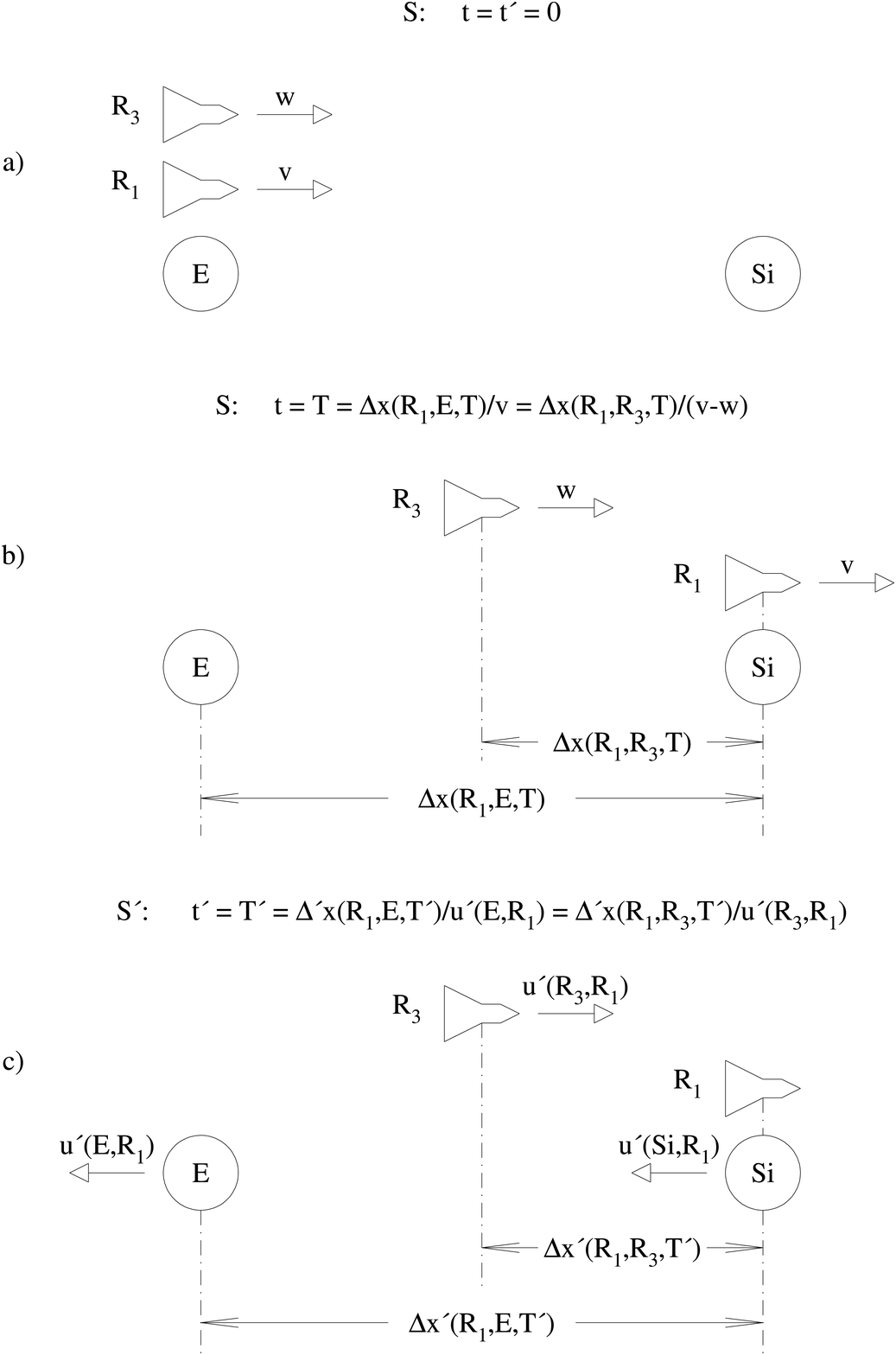}}
\caption{{\em Spatial and kinematical configurations in the outward journey of R$_1$ from the Earth (E) to
  Sirius (Si) in the twin paradox thought experiment. a) beginning of the journey in the frame S; b) end of journey
   in the frame S; c) end of journey in the rest frame, S', of  R$_1$.}}
\label{fig2}
\end{center}
\end{figure}

    \par The value of $w$ in Fig. 1a in order to yield the configuration of Fig. 1c in the rest frame of
     R$_1$ is given by (7.7) as:
    \begin{equation}
      v = u'(\RR_3,\RR_1) = \gamma(v)(v-w)
    \end{equation}
     or, transposing:
      \begin{equation}
       w = \frac{\gamma(v)-1}{\gamma(v)}v
  \end{equation}
    \par The relative velocity transformation formula (7.7) may be contrasted with the conventional
     special-relativistic parallel-velocity addition relation (PVAR) which gives instead the transformation
    of frame velocities. In the notation of Section 4, it is written:
       \begin{equation} 
     v'(\RA'',t'_\RF) = \frac{v(\RA'',t_\RF)-v(\RA',t_\RF)}{1+ \frac{v(\RA'',t_\RF)v(\RA',t_\RF)}{c^2}} 
 \end{equation}
    or, setting $v(\RA',t_\RF) \equiv v$, $v(\RA'',t_\RF) \equiv w$ and  $v'(\RA'',t'_\RF) \equiv w'$,
       \begin{equation} 
     w' = \frac{w-v}{1+ \frac{wv}{c^2}} 
 \end{equation}  
    As discussed in Refs.~\cite{JHFSTP3,JHFRECP}, this formula connects the kinematical configurations
    of a primary space-time experiment and its reciprocal, not configurations observed in  the two frames of
    the primary 
    experiment, as in (7.7). For example, setting $w = 0$ in (7.11) gives $w' = -v$ corresponding to 
     a transformation of between the world line segment of C' in S given by Eq. (6.1) to that of
    the world line segment of C in S' given by Eq. (6.4) in the reciprocal experiment.

 \SECTION{\bf{Laws of Space-Time Physics}}
      In this section some general statements concerning the physical properties of measured space 
    and time intervals as derived in Refs.~\cite{JHFSTP3,JHFRECP} and the preceding sections of the present 
    paper, are listed.
    \begin{itemize}
      \item[{\bf I}] The measured spatial interval between two objects at rest in the same reference
                     frame is independent of the reference frame in which, and epoch at which, it
                     is measured.
     \item[{\bf II}] The spatial interval between two objects in arbitary motion at a given epoch
                     is independent of the reference frame in which it is measured at the corresponding
                     epoch.
       \item[{\bf III}] Clocks which are synchronised in any reference frame (inertial or accelerated) are observed to be
                 synchronised in any other reference frame. 

      \end{itemize}  
            A `corresponding epoch' of two reference frames is defined in Section 2 above.
  \par the following definitions are introduced in describe experiments in which clocks in uniform
          motion are observed:
 \begin{itemize}
      \item[(i)] \underline{Primary Experiment} An experiment in which a clock C' at rest in 
                  the \underline{Travelling Frame} S', moving with uniform velocity $v$ in the direction
                  of the positive $x$-axis in the \underline{Base Frame} S, is compared with a similar
                  clock C at rest in S.
            
     \item[(ii)] \underline{Reciprocal Experiment} An experiment in which the clock C at rest in 
                  the \underline{Travelling Frame} S, moving with uniform velocity $v$ in the direction
                  of the negative $x'$-axis in the \underline{Base Frame} S', is compared with the clock
                  C' at rest in S'.
    \end{itemize}
                 With these definitions, the following laws concerning primary and 
                 reciprocal experiments and observed clock rates may be stated:            
     \begin{itemize}
 \item[{\bf IV}] primary and reciprocal experiments are physically independent.
\item[{\bf V}] In the primary experiment, the clock C' is seen to be running slower than C by the factor
                 $1/\gamma(v)$ by base frame observers, whereas in the travelling frame the clock C is observed
                 be running faster than C' by the factor $\gamma(v)$. 
\item[{\bf VI}] In the reciprocal experiment, the clock C is seen to be running slower than C' by the factor
                 $1/\gamma(v)$ by base frame observers, whereas in the travelling frame the clock C' is observed
                 be running faster than C by the factor $\gamma(v)$. 
 \item[{\bf VII}] The relative velocity of two objects moving parallel to the $x'$ axis in the travelling
                  frame is $\gamma(v)$ times greater than the relative velocity of the same objects in the base frame
      \end{itemize} 
             Since the concepts of primary and reciprocal experiments are not introduced in SSRT, only the first 
              parts of the laws V and VI (observation of TD from the base frame) are in agreement with the predictions
               of SSRT. The LC effect is in contradiction with I and II and RS with III. In SSRT it is assumed,
                in contradiction with VII, that velocities transform between base and travelling frames according
                 to the PVAR (7.10). The latter in fact describes instead the transformation between
              base frame configurations in a primary experiment and its reciprocal. The origin of the spurious and 
              correlated LC and RS effects of SSRT is explained in Refs.~\cite{JHFLLT,JHFSTP1,JHFCRCS,JHFACOORD,JHFUMC}.

\pagebreak

\end{document}